# Interpretable Diabetic Retinopathy Diagnosis based on Biomarker Activation Map

Pengxiao Zang, Tristan T. Hormel, Jie Wang, Yukun Guo, Steven T. Bailey, Christina J. Flaxel, David Huang, Thomas S. Hwang, and Yali Jia*

*Abstract— Objective:* **Deep learning classifiers provide the most accurate means of automatically diagnosing diabetic retinopathy (DR) based on optical coherence tomography (OCT) and its angiography (OCTA). The power of these models is attributable in part to the inclusion of hidden layers that provide the complexity required to achieve a desired task. However, hidden layers also render algorithm outputs difficult to interpret. Here we introduce a novel biomarker activation map (BAM) framework based on generative adversarial learning that allows clinicians to verify and understand classifiers' decision-making.** *Methods:* **A data set including 456 macular scans were graded as non-referable or referable DR based on current clinical standards. A DR classifier that was used to evaluate our BAM was first trained based on this data set. The BAM generation framework was designed by combing two U-shaped generators to provide meaningful interpretability to this classifier. The main generator was trained to take referable scans as input and produce an output that would be classified by the classifier as non-referable. The BAM is then constructed as the difference image between the output and input of the main generator. To ensure that the BAM only highlights classifier-utilized biomarkers an assistant generator was trained to do the opposite, producing scans that would be classified as referable by the classifier from non-referable scans.** *Results:* **The generated BAMs highlighted known pathologic features including nonperfusion area and retinal fluid.** *Conclusion/Significance:* **A fully interpretable classifier based on these highlights could help clinicians better utilize and verify automated DR diagnosis.**

## I. INTRODUCTION

D EEP learning classifiers have achieved excellent performance in several automated eye disease diagnosis tasks [1-5]. Among these tasks, diabetic retinopathy (DR) diagnosis based on optical coherence tomography (OCT) and its angiography (OCTA) play an important role in ophthalmology since DR is a leading cause of preventable blindness globally and may be asymptomatic even in the referable stages [6-9]. Therefore, an efficient and reliable diagnosis system is essential in identifying DR patients at an early stage, when the disease has the best prognosis and visual loss can be delayed or deferred [8, 9]. In addition, OCT and OCTA (which can be acquired by the same device at the same time) can provide accurate DR diagnosis based on both the standard fundus photography-derived DR severity scales [12-15] and the detection of diabetic macular edema (DME, an important DR pathology that cannot be accurately detected by fundus photography [10, 11]).

However, the high performance of current OCT/OCTA-based DR diagnosis often comes at the cost of inscrutable outputs [16-20]. The presence of hidden layers in classifier architectures renders a straightforward account of the classifier's action on inputs inaccessible and makes deep learning classifier outputs difficult to verify. In the absence of heuristic devices, deep-learning-aided DR diagnosis cannot be confirmed outside of manual grading, which largely defeats the purpose of automation. The poor interpretability can also obfuscate potential bias that could negatively affect performance in external data sets: a classifier trained only based on one data set may be biased when evaluated on the data from others if non-clinical biomarkers were utilized. These issues present a major hurdle for translating deep-learning-aided DR classifiers into the clinic [21-23].

Contemporary interpretability methods that have been used for deep-learning-aided DR diagnosis can be summarized in two categories. The first is methods which mainly focus on correlations between manually selected biomarkers and DR diagnostics [24-27]. The selected biomarkers are first segmented from OCT and OCTA and then used to train the DR classifier. However, these methods have limited DR diagnosis performance since the classifiers could not learn from the much richer feature space latent in the entire OCT/OCTA data. The second and more common interpretability methods are attention maps. They indicate the relative importance of regions of an image for classifier decision making and indicate which features were useful for the DR diagnosis. However, these methods are originally developed for non-medical image recognition tasks (e.g., dog vs. cat classification, Fig. 1) [28-30], which is distinct from DR diagnosis in several regards and

This work was supported by National Institutes of Health (R01 EY027833, R01 EY024544, P30 EY010572, T32 EY023211, UL1TR002369); Unrestricted Departmental Funding Grant and William & Mary Greve Special Scholar Award from Research to Prevent Blindness (New York, NY).

P. Zang, J. Wang, Y. Guo, D. Huang, and *Y. Jia are with Casey Eye Institute, Oregon Health & Science University, Portland, OR 97239 USA, and also with Department of Biomedical Engineering, Oregon Health & Science University, Portland, OR 97239 USA (correspondence e-mail: jiaya@ohsu.edu).

T. T. Hormel, S. T. Bailey, C. J. Flaxel, and T. S. Hwang are with Casey Eye Institute, Oregon Health & Science University, Portland, OR 97239 USA.

consequently poses many challenges in presenting clinically meaningful attention maps. In particular, unlike non-medical image classification where classes are typically identified by unique features not shared between the separate classes, in DR diagnosis different classes actually share most features (Fig. 1). Instead of containing unique identifying features, OCT/OCTA scans of non-referable DR *lack* features of referable cases. The identification of a non-referable DR case is therefore based on the absence, rather than presence, of specific pathologies.

To provide sufficient clinically meaningful interpretability for an OCT/OCTA-based DR classifier we propose a novel biomarker activation map (BAM) generation framework. The BAM generation framework is trained based on generative adversarial learning [31-34] to highlight the unique classifier-utilized biomarkers that only present in OCT/OCTA scans of referable DR. The main contributions of the present work are:

- We proposed the first interpretability method which was specifically designed for DR diagnosis based on both OCT and OCTA.
- Our design recognizes that a DR classifier should be highlight the unique biomarkers which only belong to the referable DR cases.
- We used generative adversarial learning to interpret a DR classifier instead of generating the pseudo-healthy or feature attribution maps based on the ground truth classes.
- We demonstrate that a generative adversarial learning approach can be used to provide interpretability to a DR classifier that achieved state-of-the-art performance.

## II. RELATED WORK

To achieve interpretable DR diagnosis based on OCT and OCTA, several methods were proposed to illustrate which biomarkers were utilized in the decision-making of the classifier. Most of these methods belong to one of two categories: biomarker preselection methods and attention map methods.

### A. Biomarker Preselection Methods

Biomarker preselection methods achieved interpretable DR diagnosis by using preselected and segmented biomarkers to train the classifier [24-27]. Examples include H. S. Sandhu *et al.* and M. Alam *et al.* which proposed two computer-assisted diagnostic systems for DR diagnosis based on quantified features from OCTA [25, 26]. Several DR-related biomarkers, such as foveal avascular zone size and blood vessel tortuosity and density were extracted from OCTA images to train a DR classifier. In addition, *Deep Mind* proposed a retinal disease (including DR) diagnostic system based on several pre-segmented biomarkers from OCT [27]. The interpretability issue was well addressed in these DR diagnostic tasks since clinicians can clearly know which biomarkers were used to train the classifier. However, only some biomarkers were used in these methods, which means the classifiers could not learn from the much richer feature space latent in the full OCT/OCTA data volumes.

### B. Attention Map Methods

Attention heatmaps are frequently used approaches for interpreting deep-learning-aided DR classifiers. Within this category, the gradient-based, class activation map (CAM)-based, and propagation-based methods are the most important techniques.

**Gradient-based methods** generate heatmaps based on the gradients of different convolutional layers with respect to the input [35-38]. Among these methods, Integrated Gradients which are based on multiplication between the average gradient and a linear interpolation of the input were evaluated on a DR classifier based on fundus photography [35]. To consider a more complete gradient, the FullGrad method also includes the gradient with respect to the bias term [38]. In practice, the outputs from these gradient-based methods are class-agnostic, resulting in heatmaps that are similar between different classes [39]. However, in DR diagnosis, non-referable and referable images share features, making it difficult for a gradient-based method to meaningfully distinguish pathologies from healthy tissues.

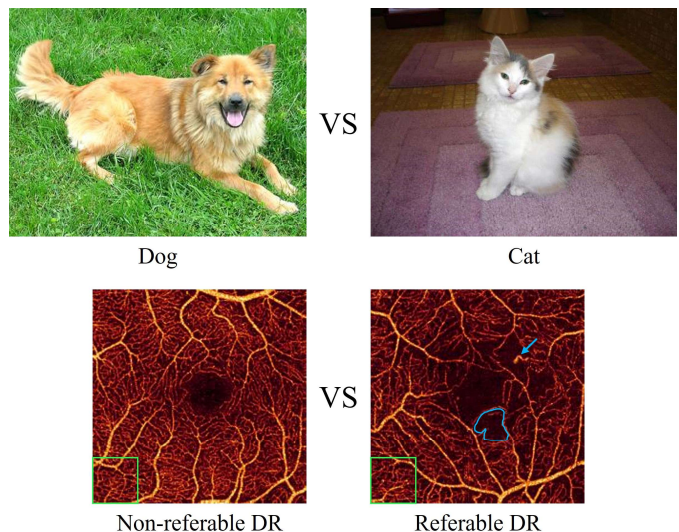

VS

Dog      Cat

VS

Non-referable DR      Referable DR

Fig. 1. Comparison between non-medical and optical coherence tomography angiography (OCTA) images. Top row: the background in dog and cat pictures can be totally different because a classifier can learn to classify each by appealing to obvious, unique features (for example, the pets' faces). This contrasts with OCTA (bottom row) of diabetic retinopathy (DR), where features are largely shared between classes (here, a non-referable and referable DR case, respectively). Additionally, only the referable DR case has unique features (DR-related biomarkers). Features found in the healthy image, for example the large vessels with surrounding small capillary are also present in the referable DR image (green rectangles). The healthy image must therefore be identified based on a lack of features associated with the referable DR image (non-perfusion area and abnormal vessels marked by blue line and arrow, respectively).

**CAM-based methods** are class-specific and are widely used in studies of deep learning DR diagnosis [40-43]. The basic CAM method combines the class-specific weight and the output of the last convolutional layer before global average pooling to produce the attention map [40]. Grad-CAM introduces the gradients of target convolutional layers to the basic CAM [41]. Grad-CAMs have been widely used in deep-learning-aided DR diagnostic studies to provide interpretability to classifiers because it is easy to implement [16, 18, 19, 20]. However, most CAM-based methods only use the top convolutional layer, which generates low-resolution heatmaps [38]. In addition, the CAMs generated on lower convolutional layers are hard to interpret due to scattered features. Clinicians still need to manually discern the biomarkers inside the coarsely highlighted regions, which is time consuming and not clinically practical.

**Propagation-based methods** [39, 44-51] mostly rely on the deep Taylor decomposition (DTD) framework [44]. In these methods, the attention map is generated by tracing the contribution of the output back to the input using back propagation through the classifier based on the DTD principle. S. Bach *et al.* proposed the Layer-wise Relevance Propagation (LRP) method, which calculates the contribution of each element in the input back propagated from the output using the DTD principle [46]. However, some of these methods are class-agnostic in practical applications [39]. To solve this issue, class-specific propagation-based methods were proposed [50, 51]. J. Gu *et al.* proposed the contrastive-LRP method in which the contributions based on non-target classes are removed on average from the heatmap [50]. B. K. Iwana *et al.* proposed the softmax-gradient-LRP in which the contribution of each non-target class is removed based on their own probability value after softmax [51]. Compared to CAMs, these class-specific LRP methods can generate higher resolution attention maps but have many fewer applications in DR diagnosis due to accuracy concerns and implementation difficulties.

Several methods which do not belong to these three attention map categories have also been proposed to interpret deep learning classifiers. These include input-modification-based methods [52-57], saliency-based methods [58-61], an activation maximization method [62], an excitation backprop method [63], and perturbation methods [64, 65]. However, these methods do not in general achieve the accuracy of gradient-, CAM-, or propagation-based methods [39].

## III. MATERIALS

In this study, we included healthy and diabetic participants representing the full spectrum of diseases, from no clinically evident retinopathy to proliferative diabetic retinopathy. One or both eyes of each participant underwent 7-field color fundus photography and an OCTA scan using a commercial 70-kHz spectral-domain OCT (SD-OCT) system (RTVue-XR Avanti, Optovue Inc) with 840-nm central wavelength. The scan depth was 1.6 mm in a 3.0 × 3.0 mm region (640 × 304 × 304 pixels) centered on the fovea. Two repeated B-frames were captured at each line-scan location. Blood flow was detected using the split-spectrum amplitude-decorrelation angiography (SSADA) algorithm [14, 66]. The OCT structural images were obtained by averaging two repeated and registered B-frames. Two continuously acquired volumetric raster scans (one x-fast scan and one y-fast scan) were registered and merged through an orthogonal registration algorithm to reduce motion artifacts [67]. In addition, the projection-resolved (PR-OCTA) algorithm was applied to all data volumes to remove flow projection artifacts in posterior layers [68, 69]. Scans with a signal strength index (SSI) lower than 50 were excluded.

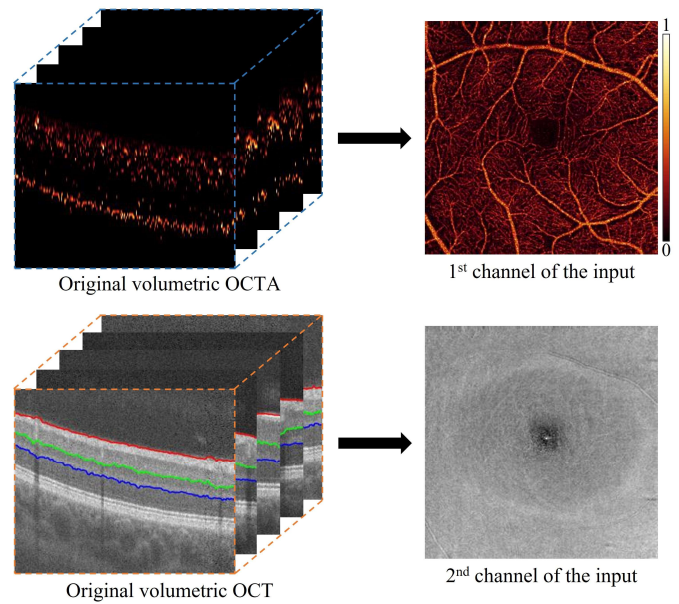

Fig. 2. Input generation. Superficial vascular complex (inner 80% from vitreous to inner plexiform layer) *en face* maximum projections were generated from the volumetric OCTA and used as the 1st channel of the input [74, 75]. An *en face* mean projection of the inner retina (vitreous through the outer plexiform layer) was generated from volumetric OCT data and used as the 2nd channel of the input. Three boundaries- vitreous / inner limiting membrane (red), inner plexiform / inner nuclear layers (green), and outer plexiform / outer nuclear layers (blue)- were segmented for the generation process in both the first and second channels.

A masked trained retina specialist (Thomas S. Hwang) graded the photographs based on the Early Treatment of Diabetic Retinopathy Study (ETDRS) scale [70, 71] using 7-field color fundus photographs. The presence of diabetic macular edema (DME) was determined using the central subfield thickness from structural OCT based on the DRCR.net standard [8]. We defined non-referrable DR as ETDRS level better than 35 and without DME, and referrable DR as ETDRS level 35 or worse, or any ETDRS score with DME [9]. The participants were enrolled after informed consent in accordance with an Institutional Review Board (IRB # 16932) approved protocol. The study complied with the Declaration of Helsinki and the Health Insurance Portability and Accountability Act.

To generate the input data set for the referable DR classifier, the following retinal layer boundaries were automatically segmented using commercial software in the SD-OCT system (Avanti RTVue-XR, Optovue Inc): the vitreous / inner limiting membrane (ILM), inner plexiform layer (IPL) / inner nuclear layer (INL), and the outer plexiform layer (OPL) / outer nuclear layer (ONL) (Fig. 2). In addition, the automated layer segmentation was manually corrected by graders using custom COOL-ART software in cases where pathology caused

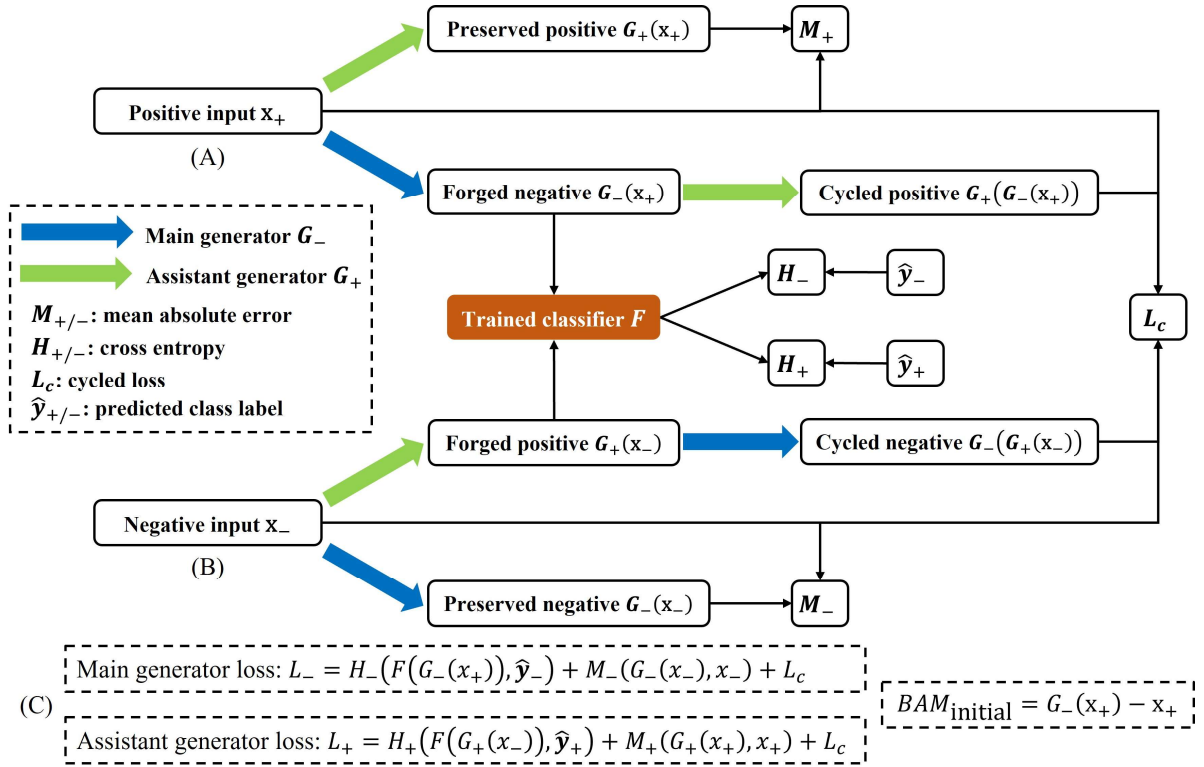

**Preserved positive** $G_+(\mathbf{x}_+)$ → $M_+$

**Positive input** $\mathbf{x}_+$

(A)

**Forged negative** $G_-(\mathbf{x}_+)$ → **Cycled positive** $G_+(G_-(\mathbf{x}_+))$

— **Main generator** $G_-$
— **Assistant generator** $G_+$

$M_{+/-}$: mean absolute error
$H_{+/-}$: cross entropy
$L_c$: cycled loss
$\hat{y}_{+/-}$: predicted class label

**Trained classifier** $F$ → $H_-$ ← $\hat{y}_-$
→ $H_+$ ← $\hat{y}_+$

$L_c$

**Forged positive** $G_+(\mathbf{x}_-)$ → **Cycled negative** $G_-(G_+(\mathbf{x}_-))$

**Negative input** $\mathbf{x}_-$

(B)

**Preserved negative** $G_-(\mathbf{x}_-)$ → $M_-$

(C) Main generator loss: $L_- = H_-\big(F(G_-(\mathbf{x}_+)), \hat{y}_-\big) + M_-(G_-(\mathbf{x}_-), \mathbf{x}_-) + L_c$

Assistant generator loss: $L_+ = H_+\big(F(G_+(\mathbf{x}_-)), \hat{y}_+\big) + M_+(G_+(\mathbf{x}_+), \mathbf{x}_+) + L_c$

$BAM_{\text{initial}} = G_-(\mathbf{x}_+) - \mathbf{x}_+$

Fig. 3. BAM generation framework architecture and training process. Two generators were trained to produce the biomarker activation maps (BAMs). (A) Positive class inputs are acted on by the main and assistant generators (blue and green arrows, respectively). The main generator was trained to produce output images that would be classified as the negative class by the classifier. The assistant generator performs the inverse task, producing outputs that the classifier would diagnose as the positive class. (B) Training from negative inputs is symmetric, with the labels switching roles. (C) Training of the main and assistant generators occurred simultaneously. This scheme prevents the main generator from overfitting the negative class inputs by producing unnecessary changes to the inputs.

segmentation errors in the commercial software [72].

For each case, a two-channel input was generated based on the segmented boundaries (Fig. 2). The first input consists of the *en face* OCTA image was generated using maximum projection of the superficial vascular complex (SVC), defined as the inner 80% of the ganglion cell complex (GCC), which included all structures between the ILM and IPL/INL border [74, 75] (Fig. 2) [73]. *En face* structural OCT images were generated through average projection (Vitreous/ILM to OPL/ONL) and used as the 2nd channel of each input (Fig. 2).

## IV. METHODS

To provide meaningful interpretability to a DR classifier, our BAM generation framework was trained to learn which biomarkers were important for classifier decision making. In training, we combined two generators (a main and assistant) to learn the necessary changes (classifier-utilized biomarkers) which would change classifier decision making. In this study, the positive and negative decision making of the classifier corresponded to referable and non-referable DR, respectively. The main generator was trained to forge a negative output by adding changes to a positive input. To reduce unnecessary changes made by the main generator, inspired by cycle-consistency generative adversarial learning [34], the assistant generator was trained to do the opposite. However, the BAM was generated as the difference image between the output and input of the main generator, only. All the biomarkers highlighted in the BAM were the classifier-utilized biomarkers, which could be textures, artifacts, shadows, etc. that were learned and utilized by the classifier in the decision making. To be clear, the classifier-utilized biomarkers could be different from the clinical biomarkers which mainly were the pathologies correlated to the selected disease (referable DR in this study).

### A. Training

We consider a DR classifier $F$ trained to predict positive and negative class labels $\hat{y}$ from input data $\mathbf{x}$ according to $\hat{y} = F(\mathbf{x})$. The positive and negative class $y_+$ and $y_-$ indicated referable and non-referable DR, respectively. We note that, in general, the predicted class label $\hat{y}$ is not identical to the true class labels, $y$, since most classifiers are not perfect; however, in this work we are primarily concerned with classifier outputs, not the ground truth classifications. Accordingly, we define two input classes $\mathbf{x}_+$ and $\mathbf{x}_-$ according to $F(\mathbf{x}_{+/-}) = \hat{y}_{+/-}$, *i.e.* $\mathbf{x}_+$ corresponds to data that was predicted by the classifier to belong to the positive class (i.e. referable DR), and $\mathbf{x}_-$ to the negative class. In the BAM framework we seek to train a main generator to transform data so that it is always classified as negative by the classifier (Fig. 3(A)); that is, we seek a generator $G_-$ such that $F(G_-(\mathbf{x})) = \hat{y}_-$. In the case that the input data was originally classified as positive, i.e. $\mathbf{x} = \mathbf{x}_+$, this creates "forged data", in which the target output classification $\hat{y}_-$ differs from the classification of the input for which $F(\mathbf{x}_+) = \hat{y}_+$. If, alternatively, $G_-$ operates on data $\mathbf{x}_-$ already predicted to belong to the negative class, the desired output classification matches the input, i.e. $F(G_-(\mathbf{x}_-)) = \hat{y}_-$, creating "preserved data." Both forged and preserved data were used during the training process to calculate loss by comparing to their corresponding ground truths. Specifically, the cross-entropy loss $H_-$ between the classifier prediction on the forged data $F(G_-(\mathbf{x}_+))$ and the desired prediction $\hat{y}_-$ was used to train the generator to produce data resembling the desired class. In addition, to prevent large changes to the main generator output, the mean absolute error loss $M_-$ between the raw input $\mathbf{x}_-$ and the preserved data $G_-(\mathbf{x}_-)$ was included in the loss.

However, simply optimizing over forged and preserved data can lead to overfitting, in which the main generator learns to modify features that were not utilized by the classifier $F$ (e.g., shared features between $\mathbf{x}_+$ and $\mathbf{x}_-$) in order to achieve the desired output label $\hat{y}_-$. To ensure that the main generator only learns to remove relevant features we also simultaneously trained an assistant generator $G_+$ that performs the inverse task; that is, we desire the trained assistant generator to produce $F(G_+(\mathbf{x})) = \hat{y}_+$ (Fig. 3(B)). Note that, like the main generator, this produces both preserved and forged data, since the assistant generator also acts on both $\mathbf{x}_+$ and $\mathbf{x}_-$. The assistant generator is used in conjunction with the main generator to produce "cycled data" $G_-(G_+(\mathbf{x}_-))$ and $G_+(G_-(\mathbf{x}_+))$ created by allowing the main and assistant generator to operate on data forged by the other. The cycled loss, defined as the mean absolute errors between the original and cycled data

$$L_c = \frac{1}{N_-}\sum_i \left|\mathbf{x}_{-,i} - G_-\left(G_+(\mathbf{x}_{-,i})\right)\right| +$$
$$\frac{1}{N_+}\sum_j \left|\mathbf{x}_{+,j} - G_+\left(G_-(\mathbf{x}_{+,j})\right)\right|, \quad (1)$$

where $N_+$ and $N_-$ are the pixel number of positively and negatively classified images, respectively, can then be included in the overall loss function in order to ensure that only features utilized by the classifier $F$ are modified. The overall loss for each generator is then given by the sum of the cross-entropy loss between forged labels and predicted labels, the mean absolute error loss between preserved data and input data, and the cycled loss:

$$L_- = H_-\left(F(G_-(x_+)), \hat{y}_-\right) + M_-(G_-(x_-),\ x_-) + L_c$$
$$L_+ = H_+\left(F(G_+(x_-)), \hat{y}_+\right) + M_+(G_+(x_+),\ x_+) + L_c \quad (2)$$

### B. Generator Architecture

Both the main and assistant generators were constructed based on a customized U-shape residual convolutional neural network [76] (Fig. 4). The output is calculated as the sum of input and Tanh output since both generators are trained to only change necessary biomarkers that are utilized by the classifier. To ensure the generator output has the same value range as the input, clipping was used after the summation (the values higher or lower than the original maximum or minimum of the input will be set to the maximum or minimum values, respectively). In addition, zero initialization is used for the last convolutional layer to ensure the initial BAM was a zero matrix before the training. This initialization strategy could avoid changes to biomarkers which were not utilized by the classifier in the beginning of the training.

### C. Model Selection and BAM Generation

After the training, the final model (including both trainable parameters and hyper-parameters) for BAM generation was selected based on the validation loss $L_-$ of the main generator. Loss from the assistant generator was not considered because, unlike the main generator which only needs to remove the unique biomarkers (pathologies) at specific locations, there is no a priori reason for the assistant generator to add pathology-like features at a particular location. This resulted in high variability in assistant generator output, which would make model selection based on assistant generator loss unreliable.

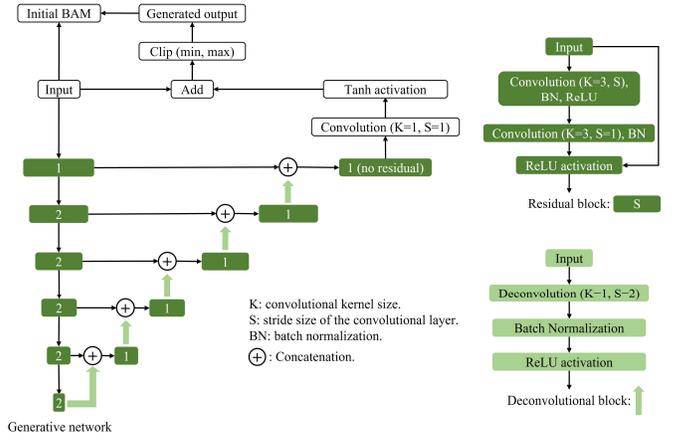

Fig. 4. Detailed architecture of the main generator. The dark green patches and pale green arrows represent the residual and deconvolutional block, respectively. The number in the dark green patch is the stride size. The number of blocks can be adjusted based on the input. The architecture of the assistant generator is identical.

The initial BAM is calculated between the output and input of the main generator. Since this is a difference image it can have both positive and negative pixel values. The absolute difference between these values represents the overall contribution each biomarker made to the classification. Alternatively, positive/negative values indicate regions in which pixel values in the output were increased/reduced relative to the input for the classifier to produce a negative (non-referable DR) classification. By keeping track of these sign differences, we can understand more about how a DR classifier understands different biomarkers. Accordingly, the output of our framework is two processed BAMs, with the first obtained by measuring the absolute value of all differences

$$BAM_{abs} = f_g(|G_-(\mathbf{x}_+) - \mathbf{x}_+|), \quad (3)$$

while the second is generated by separating positive and negative values

$$BAM_{+/-} = f_g\left(\text{ReLU}(G_-(\mathbf{x}_+) - \mathbf{x}_+)\right) -$$
$$f_g\left(\text{ReLU}(\mathbf{x}_+ - G_-(\mathbf{x}_+))\right), \quad (4)$$

where $f_g$ is a Gaussian filter and ReLU is the activation function which only preserves positive values [77, 78]. The $BAM_{abs}$ then indicates the overall contribution of each biomarker to classifier decision making, while the $BAM_{+/-}$ indicates how different biomarkers were learned by the classifier.

### D. Implementation Details

To evaluate our BAM generation framework, a classifier that took en face OCT and OCTA data as inputs to diagnose referable DR was constructed based on a VGG19 architecture [78] with batch normalization and only one fully connected layer. Two classifier-utilized DR biomarkers- non-perfusion area (NPA) and fluids- were used to evaluate the OCTA and OCT channels of the generated BAMs, respectively. The DR classifier was only used to evaluate our BAMs. The development of this classifier was not a part of our BAM generation framework.

Before the evaluations, 60%, 20%, and 20% of the data were split for training, validation, and testing, respectively. Care was taken to ensure data from the same subjects are only included in one of either the training, validation, or testing data sets. The BAM framework was trained, validated, and evaluated respectively based on the same data set of the DR classifier. Two stochastic gradient descent optimizers with Nesterov momentum (momentum = 0.9) were used simultaneously to train the generators. Hyperparameters during training included a batch size of 3, 500 training epochs, and a learning rate of 0.0005 used for all the training steps. The trained main generator with lowest validation loss was selected for the final evaluation. In the evaluation, the binary maps for NPA and fluids were generated by two deep-learning-aided segmentation methods previously designed by our group, respectively [84, 85]. For qualitative analysis, we used perceptually uniform color maps to illustrate the BAMs [80]; compared to the traditional Jet colormap, perceptually uniform color maps have even color gradients that can reduce visual distortions causing feature loss or the appearance of false features [81]. For quantitative analysis, the F1-socre, intersection over union (IoU), precision, and recall were calculated between each channel of the generated BAMs and segmented DR biomarkers (NPA and fluid).

This study was implemented in TensorFlow version 2.6.0 on Ubuntu 20.04 server. The server has an Intel(R) Xeon(R) Gold 6254 CPU @ 3.10GHz ×2, 512.0 GB RAM and four NVIDIA RTX 3090 GPUs. But only one GPU was used in this study. The average training time for each epoch was 15 seconds.

### E. Sanity Checks

To assess if our BAM was correlated with the interpreted DR classifier two sanity checks were performed [82]. First, we performed model parameter and data randomization tests. In the model parameter randomization test DR classifier parameters were divided into six parts based on the five max pooling layers. The parameters in each of the six parts were randomized in two ways. In cascading randomization we randomized the parameters from the top part of the trained DR classifier (after last max pooling) successively all the way to the bottom part (before first max pooling). In the independent randomization, the parameters in each part were randomized independently. All the parameter randomizations created 11 different models. In the data randomization test, a model with the same architecture of the DR classifier was trained based on randomized labels. The model training was stopped after the training accuracy reached 95%. The generated BAMs of these 12 models were compared with the BAMs generated based on the original DR classifier. For quantitative comparison, we calculated the spearman rank correlation, the structural similarity index (SSIM), and the Pearson correlation of the histogram of gradients (HOGs) between the $BAM_{+/-}$ generated on these models and the original classifier.

## V. RESULTS

TABLE I
DATA DISTRIBUTION

| Severity | Number | Age, mean (SD), years | Female, % |
|---|---|---|---|
| Non-referable DR | 199 | 48.8 (14.6) | 50.8% |
| Referable DR | 257 | 58.4 (12.1) | 49.0% |

DR = diabetic retinopathy.

We recruited and examined 50 healthy participants and 305 patients with diabetes. After the DR severity grading, 199 non-referable and 257 referable DR inputs were used to train, validate, and evaluate the DR classifier and our framework (Table I). In the evaluation, the classifier achieved an area under the receiver operating characteristic curve (AUC) of 0.97 and a quadratic-weighted kappa of 0.85, which is on par with the performance of ophthalmologists and therefore adequate to evaluate our BAM framework [83].

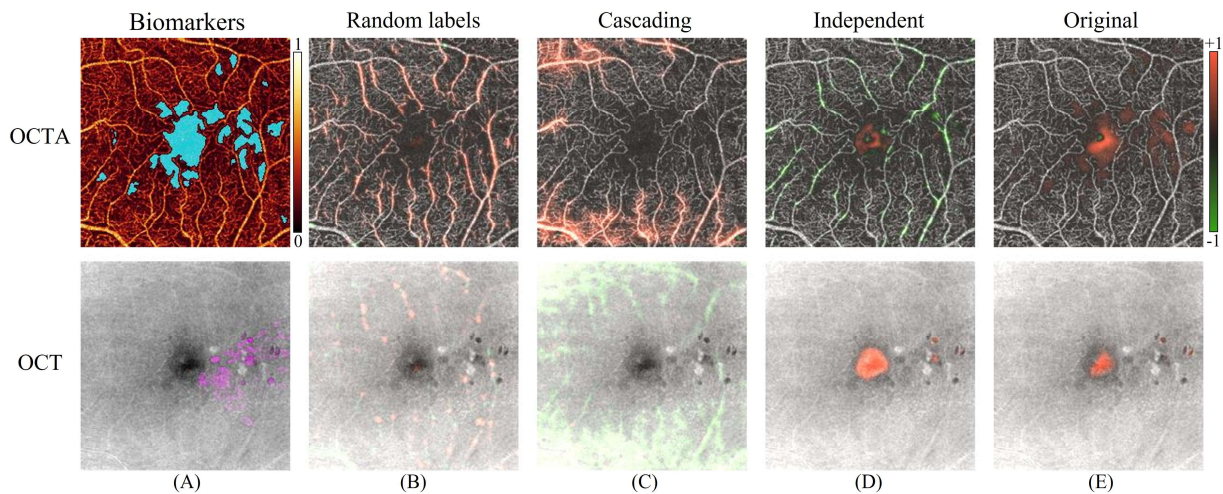

Fig. 5. BAMs generated in the sanity checks which showed our BAM was sensitive to the interpretability changes between different randomized models. (A) Segmented non-perfusion areas and fluids. (B) The $BAM_{+/-}$ of the model based on randomized labels. (C) The $BAM_{+/-}$ of the model based on cascading parameter randomization. (D) The $BAM_{+/-}$ of the model based on independent parameter randomization. (E) The $BAM_{+/-}$ generated based on the original DR classifier.

TABLE II
QUANTITATIVE SANITY CHECKS

| Models | Spearman rank correlation | | SSIM | | HOGs | |
|---|---|---|---|---|---|---|
| | OCTA | OCT | OCTA | OCT | OCTA | OCT |
| Random labels | -0.11 ± 0.17 | 0.22 ± 0.22 | 0.13 ± 0.07 | 0.75 ± 0.11 | 0.07 ± 0.06 | 0.01 ± 0.09 |
| Cascading randomization | -0.21 ± 0.06 | 0.11 ± 0.03 | 0.18 ± 0.10 | 0.24 ± 0.11 | 0.00 ± 0.06 | -0.09 ± 0.06 |
| Independent randomization | 0.45 ± 0.14 | 0.82 ± 0.11 | 0.30 ± 0.12 | 0.91 ± 0.12 | 0.05 ± 0.05 | 0.54 ± 0.15 |

OCT = optical coherence tomography, OCTA = optical coherence tomography angiography, SSIM = structural similarity index, HOGs = Pearson correlation of the histogram of gradients.

## A. Sanity Checks

To firstly demonstrate that our BAM could correctly learn the interpretability of the DR classifier, two tests were performed based on randomized parameters and labels, respectively [82]. Most models in the parameter randomization test predicted all the data as the same class (either non-referable or referable DR), which means no BAMs were generated for these models since the training of our framework needed data predicted as both classes. In the cascading randomization, only the model with randomized parameters after the 4th max pooling layer had predictions for both classes. In the independent randomization, only the model with randomized parameters before the first max pooling layer had predictions for both classes. Therefore, these two models were used to represent the cascading and independent parameter randomizations, respectively. The three BAMs generated in both parameter and label randomization tests showed large differences compared to the original BAMs (Fig. 5 and Table II), which shows our BAMs are sensitive to potential interpretability changes of the classifier. The two models based on randomized labels and cascading parameter randomization highlighted totally different regions compared to the original BAMs. The highlighted regions in the model based on independent randomization had some overlaps with the original BAMs. But the differences between these two BAMs were still large and clear.

## B. Qualitative Analysis

To demonstrate the utility of the proposed BAM framework we consider an eye correctly classified as referable DR by the DR classifier (Fig. 6). Compared to the clinical DR biomarkers (Fig. 6(C) and 6(H)), our BAMs highlighted similar regions (Fig. 6(D) and 6(I)), demonstrating that BAMs can improve interpretation of the DR classifier result. Specifically, the $BAM_{abs}$ output indicates that the classifier focused on important pathologic features which (NPA and retinal fluid) [84, 85] in decision making. Additionally, the $BAM_{+/-}$ indicates that the pathological NPA was correctly differentiated from the foveal avascular zone, which is avascular but not pathological. In non-referable eyes, the pixel values near fovea should span a wide range corresponding to- from highest to lowest- larger vessels around fovea, small capillary structure, and the foveal avascular zone. These pixel populations are compressed in an *en face* OCTA angiograms of a referable DR eye (Fig. 6(A)). By adding positive values (white dots) around fovea and negative values in the fovea, the main generator expanded the range of pixel values to craft an image that looked like a non-referable eye to the classifier (Fig. 6(B)). The structure of the generated image shows that the classifier learned the anatomical structure near fovea in a normal eye. We also noticed that, with the exception of NPA and fluid, other DR-related biomarkers (such as microaneurysms, Fig. 6(I)) were ignored by the BAMs, which means these biomarkers were not utilized by the classifier in decision making.

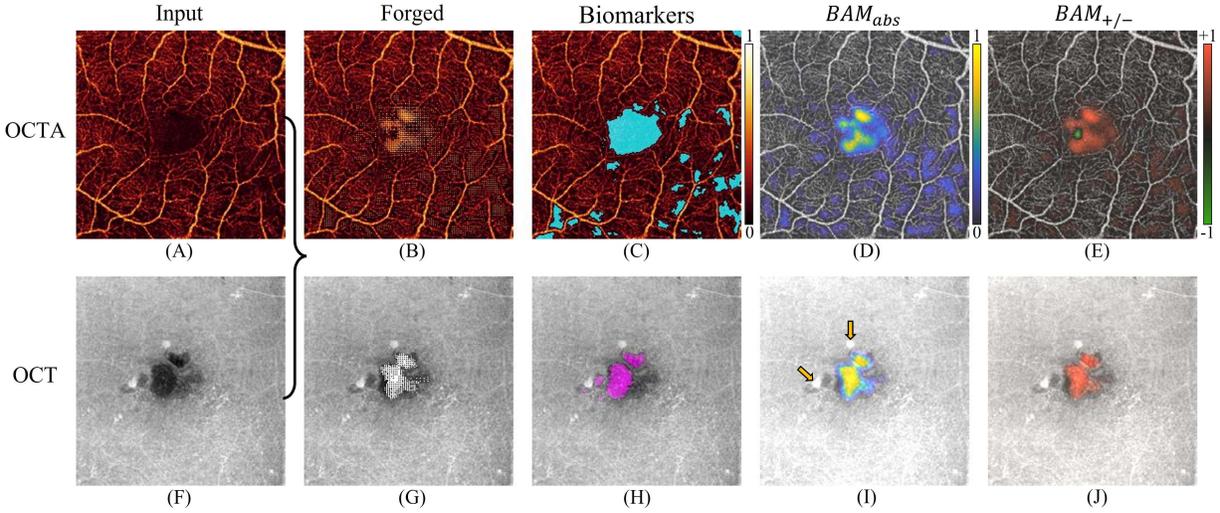

Fig. 6. BAMs for a correctly predicted referable DR scan. (A) Superficial vascular complex *en face* maximum projection, which is the first input channel for the BAM framework. (B) The forged main generator output image for this channel, which should be classified as non-referable DR by the classifier. Compared to (A), positive white dots and negative dark regions were added by the main generator. (C) Segmented non-perfusion area (NPA), which is an important DR-related biomarker (marked by cyan), based on a previously reported deep learning method [84]. (D) The $BAM_{abs}$ is the absolute difference between (B) and (A) after Gaussian filtering (Eq. 3). The highlighted areas are similar to the segmented NPA in (C). (E) The $BAM_{+/-}$ is the (non-absolute) differences between (B) and (A) after Gaussian filtering (Eq. 4). Red highlights pathological non-perfusion area while the foveal avascular zone (highlighted by green), which is not pathological, was identified as a separate feature by the classifier network. (F) *En face* mean projection over the inner retina, which is the second input channel to the classifier. (G) Main generator output for this channel which should be classified as non-referable DR by the classifier. Compared to (F), positive white dots were added by the main generator. (H) The inner mean projection of the segmented fluid (an important DR-related biomarker, marked by magenta) based on a previously reported deep learning method [85]. (I) The $BAM_{abs}$ is the absolute difference between (G) and (F) after Gaussian filtering (Eq. 3). The highlighted areas resemble the fluid regions in (H). However, the microaneurysms (hyperreflective spots marked by orange arrows) were not highlighted, which means this biomarker was not utilized by the classifier. (J) The $BAM_{+/-}$ is the difference between (G) and (F) after Gaussian filtering (Eq. 4). The red highlighted areas also focus on fluids, and no green highlighted area is shown, indicating that the network did not learn separate fluid features. This is anatomically accurate, since unlike NPA (which is non-pathologic in the foveal avascular zone) all retinal fluid is pathologic.

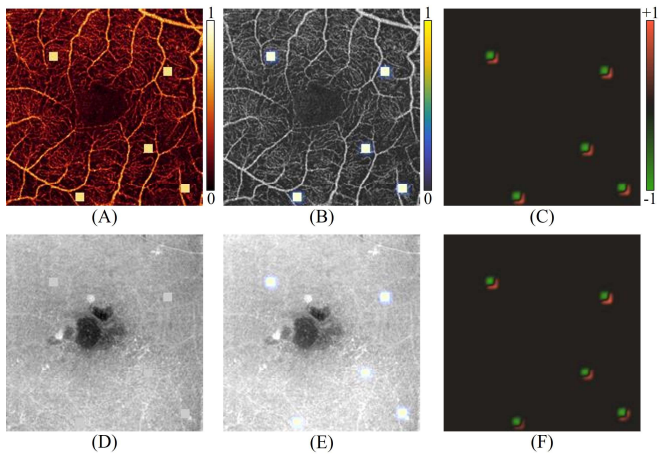

Fig. 7. BAMs for a correctly predicted referable DR scan based on a biased classifier. (A) Superficial vascular complex *en face* maximum projection with added artifacts. (B) OCTA channel of the $BAM_{abs}$ which only highlighted the classifier-utilized artifacts. (C) OCTA channel of the $BAM_{+/-}$ without coverage of input. (D) *En face* mean projection over the inner retina with added artifacts. (E) The OCT channel of $BAM_{abs}$ which only highlighted the classifier-utilized artifacts. (F) OCT channel of the $BAM_{+/-}$ without coverage of input.

To demonstrate how our BAM could help detect biased classifiers 5 rectangular artifacts were added to both the OCTA and OCT channels of all the referable DR scans. In the training, the classifier was forced to utilize the artifacts to make predictions, thereby introducing bias into the classifier output. Our BAM network trained on this biased classifier was able to successfully highlight the artifactual features (Fig. 7). The highlighted artifacts (Fig. 7(B) and 7(E)) show that the clinical biomarkers were not utilized by the biased classifier, despite it reaching 100% accuracy. The $BAM_{+/-}$ (Fig. 7(C) and 7(F)) reveals how the biased classifier utilized the artifacts. In addition, these artifacts could be much less obvious in a real-world application (such as Fig. 7(D)).

To demonstrate the advantages of the BAM generation framework, we compared its output to gradient-based [35], propagation-based [51], and CAM-based methods [40] (Fig. 8). Compared to the attention maps generated by these methods in referable DR scans, the BAMs showed sharper distinctions between significant and insignificant regions for decision making, highlighted features at a higher resolution, and indicated that the classifier could distinguish different features. In addition, the BAM framework could separately highlight the important features in *en face* OCTA and structural OCT rather than blending them together. This improves interpretability since the features the network is trying to learn do not necessarily overlap in the separate channels. For example, NPA does not always overlap with diabetic macular edema. Especially for graders reviewing the images, if structural OCT and OCTA features are not separated it may be unclear if healthy regions in one channel are being misinterpreted as pathologic, or if the pathology is in the other channel. However, gradient-based, CAM-based, and propagation-based methods all highlighted similar regions between two different channels (Fig. 8). Several NPAs were ignored in the OCTA channel. In addition, small retinal fluids, which were highlighted by our BAMs, were also ignored by these three attention maps (marked by orange arrows in Fig. 8(B)).

In non-referable DR scans our BAM still highlighted a sharp foveola because due to the classifier needing to segment non-pathological NPA in the foveal avascular zone. Since this feature exists in both referable and non-referable classes, it is also highlighted in the non-referable case by the BAM (Fig. 8(C)). The gradient-based and CAM-based maps of the non-referable class highlighted similar areas compared to the maps for the referable DR class, which means their highlighting was incorrect in non-referable DR class. The propagation-based method correctly highlighted the surrounding areas of foveola, but the highlighting was inaccurate since areas without highlighting were much larger than the foveola.

## C. Quantitative Analysis

To compare our BAM with the other three attention maps quantitatively, the F1-score, IoU, precision and recall were calculated between the segmented biomarkers and binary masks of each attention map (Table III). The binary mask of each map was generated using threshold $mean + std \times h$ on all positive values. For each attention method, the threshold $h$ was selected based on the highest average F1-score. The OCTA channels of each attention map were compared with segmented NPAs. On this channel, our BAM achieved significantly higher performance than the other three attention maps (Table III). On the OCT channels, which were compared with segmented fluids, our BAM still achieved higher performance based on most measurements. Only recall for fluids was lower than the other methods. But the higher precision and lower recall of the BAM actually demonstrate that our method mostly focused on the part of the DR biomarkers which were utilized by the classifier while ignored the healthy tissues. The lower precision and higher recall of these established attention maps can be

TABLE III
QUANTITATIVE COMPARISON

| Methods | Inference time s/scan | F1-score | | IoU | | Precision | | Recall | |
|---|---|---|---|---|---|---|---|---|---|
| | | NPAs | Fluids | NPAs | Fluids | NPAs | Fluids | NPAs | Fluids |
| Gradient | 0.10 | $0.39 \pm 0.07$ | $0.10 \pm 0.15$ | $0.24 \pm 0.05$ | $0.06 \pm 0.10$ | $0.40 \pm 0.08$ | $0.07 \pm 0.11$ | $0.41 \pm 0.14$ | $0.54 \pm 0.34$ |
| Propagation | 0.07 | $0.42 \pm 0.08$ | $0.11 \pm 0.17$ | $0.27 \pm 0.06$ | $0.07 \pm 0.12$ | $0.40 \pm 0.11$ | $0.08 \pm 0.15$ | $0.48 \pm 0.12$ | $\mathbf{0.62 \pm 0.35}$ |
| CAM | 0.05 | $0.44 \pm 0.09$ | $0.11 \pm 0.17$ | $0.29 \pm 0.07$ | $0.07 \pm 0.12$ | $0.46 \pm 0.10$ | $0.08 \pm 0.13$ | $0.46 \pm 0.15$ | $0.54 \pm 0.36$ |
| BAM | 0.07 | $\mathbf{0.63 \pm 0.09}$ | $\mathbf{0.13 \pm 0.20}$ | $\mathbf{0.47 \pm 0.09}$ | $\mathbf{0.09 \pm 0.14}$ | $\mathbf{0.58 \pm 0.15}$ | $\mathbf{0.20 \pm 0.31}$ | $\mathbf{0.72 \pm 0.07}$ | $0.21 \pm 0.29$ |

NPAs = non-perfusion areas, CAM = class activation map, BAM = biomarker activation map, IoU = intersection over union.

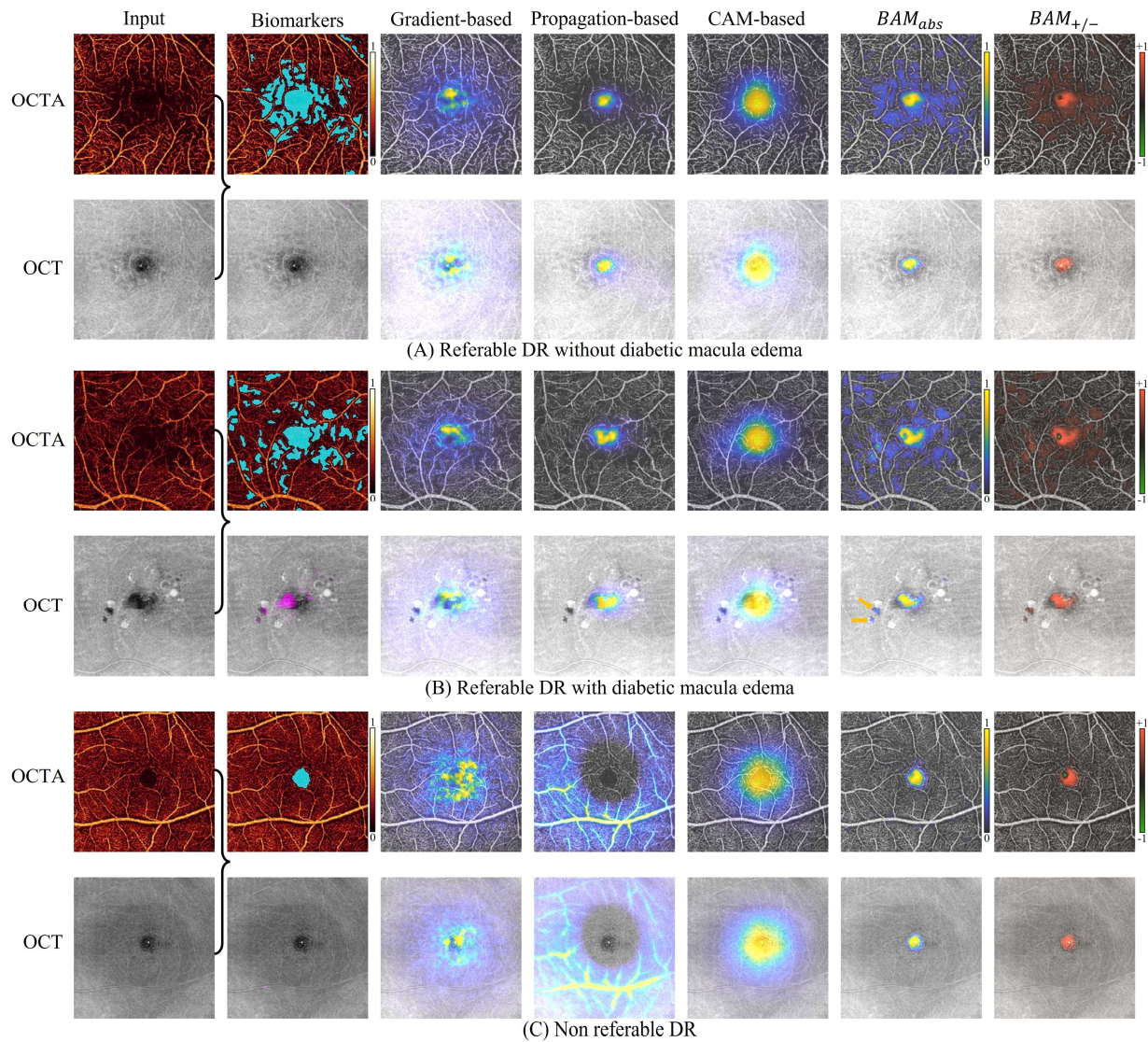

|  | Input | Biomarkers | Gradient-based | Propagation-based | CAM-based | $BAM_{abs}$ | $BAM_{+/-}$ |

(A) Referable DR without diabetic macula edema

(B) Referable DR with diabetic macula edema

(C) Non referable DR

Fig. 8. Comparison between the BAM generation framework and three other prominent attention maps (gradient, propagation, and class activation) [35, 41, 51]. The biomarkers column shows non-perfusion area (NPA, marked by cyan) in OCTA channel and retinal fluid (marked by magenta) in the structural OCT channel, respectively, both segmented using previously reported deep learning methods [84, 85]. Compared to the other three attention maps, our BAMs accurately highlight each classifier-selected biomarker at higher resolution and highlighted just the classifier-selected biomarkers. In addition, the normal tissues (such as vessels between NPAs) that were not selected by the classifier were not highlighted by our BAMs. (A) Results based on a referable DR case without diabetic macular edema (DME). (B) Results based on a referable DR case with DME. Small fluids were sharply highlighted by the BAMs (marked by orange arrows). (C) Results based on a non-referable DR case. The BAM highlighted a sharp foveola because our BAM would always highlight the areas which were learned as referable DR biomarkers by the classifier. The gradient-based and CAM-based maps of the non-referable class highlighted similar areas compared to the maps of referable DR class, which means their highlighting was incorrect in non-referable DR class. The propagation-based method correctly highlighted the surrounding areas of foveola, but the highlighting was inaccurate since areas without highlighting were much larger than the foveola.

attributed to the fact that they highlighted a large area which included more healthy tissues than DR biomarkers, which is not clinically meaningful. In addition, all four attention maps achieved higher performance on the OCTA channel than the OCT channel, which means the classifier was more focused on the NPAs rather than fluids.

Based on the generated BAMs (Fig. 6 and 8, Table III), the interpretability of the DR classifier can be summarized as follows. First, only parts of the NPA and fluid regions were utilized by the classifier (Fig. 6 and 8). Other DR-related biomarkers (such as microaneurysms, Fig. 6(I)) were not utilized by the classifier. Second, pathological NPAs were correctly differentiated from the foveal avascular zone by the classifier (Fig. 6(E)). Third, the classifier utilized the foveola in

OCT channel even though this dark region was not caused by fluids (BAMs in Fig. 8(A)). Lastly, more NPAs were utilized than fluid areas in the decision making of the classifier (Fig. 6 and 8, Table III).

### D. Biomarkers Analysis

As sanity checks, qualitative and quantitative analyses above demonstrated that our BAM could provide sufficient interpretability to a DR classifier based on OCT/OCTA. To find all the biomarkers which could contribute to referable DR diagnosis, the BAMs were also generated for two classifiers (0.75 and 0.71 kappas) trained based only on the OCTA and OCT scans, respectively (Fig. 9). In the OCTA scans, compared to the DR classifier trained with two channel inputs, the BAMs

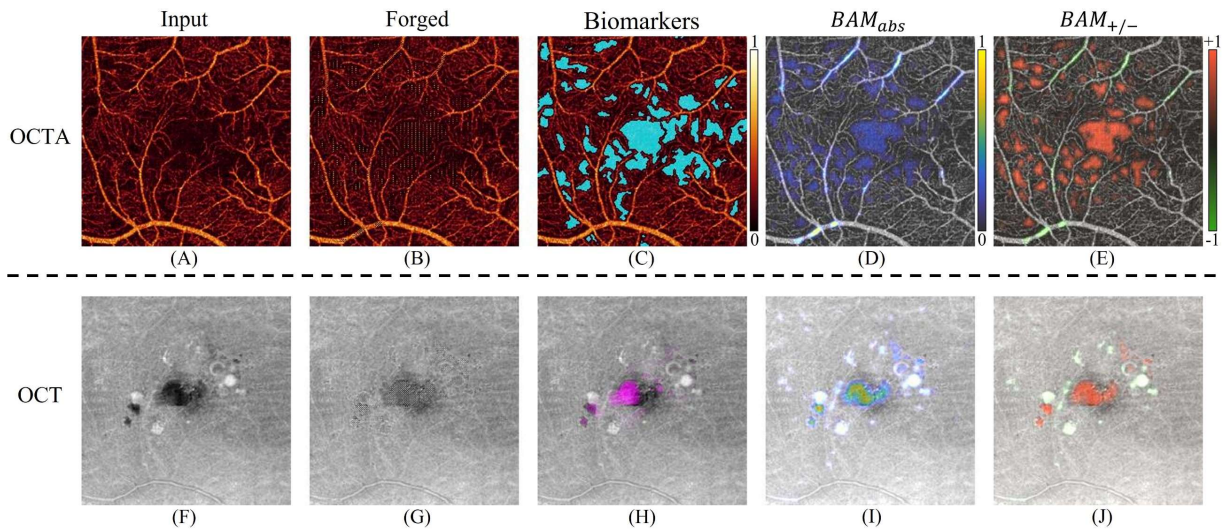

Fig. 9. BAMs generated for two DR classifiers trained based only on the OCTA and OCT scans, respectively. (A) Superficial vascular complex *en face* maximum projection of OCTA. (B) The forged main generator output. (C) Segmented non-perfusion area (NPA) based on a previously reported deep learning method [84]. (D) The $BAM_{abs}$ is the absolute difference between (B) and (A) after Gaussian filtering (Eq. 3). The highlighted areas are similar to the segmented NPA in (C). (E) The $BAM_{+/-}$ is the (non-absolute) differences between (B) and (A) after Gaussian filtering (Eq. 4) Except to the NPAs highlighted by red (positive values), parts of the vessels of high intensities were highlighted by green (negative values). (F) *En face* mean projection over the inner retina of OCT. (G) The forged main generator output. Hypo-reflective fluids and hyperreflective spots in (F) were both changed typical reflectivity values. (H) The inner mean projection of the segmented fluid based on a previously reported deep learning method [85]. (I) The $BAM_{abs}$ is the absolute difference between (G) and (F) after Gaussian filtering (Eq. 3). (J) The $BAM_{+/-}$ is the difference between (G) and (F) after Gaussian filtering (Eq. 4). The red highlighted areas focus on fluids, and green areas focus on abnormal hyperreflective spots.

also highlighted most of the NPAs but with basically equal intensity (Fig. 9(D)), which means prediction contributions from the NPAs outside the foveola were improved when OCTA was the only input. In addition, small parts of the vessels with higher intensities were also highlighted (Fig. 9(E)). In the OCT scans, compared to the OCT channel of Fig. 7(B), all the fluid and other hyperreflective regions were highlighted by the BAMs (Fig. 9(I)). In addition, hyperreflective spots which were not highlighted before were also highlighted this time (Fig. 9(J)) because referable DR could not be detected only based on fluid. In summary, the NPAs, fluid, and abnormal hyperreflective spots (could be exudation, calcification, and microaneurysm) could all contribute to the deep-learning-aided

DR diagnosis, which is consistent with clinical findings. In addition, some vessel parts with high intensity, hypo-reflective areas without fluids, and hyperreflective spots without pathologies were also highlighted by the BAMs. The highlighting of these non-clinical biomarkers may be caused by the imperfect classifiers or potential correlations that have not been found.

### E. Ablation Experiments

In the proposed framework, both $M_{+/-}$ and $L_c$ losses were used to ensure that the BAM framework only highlighted the classifier-utilized biomarkers. But the use of these two losses also reduced the computational efficiency. To explore its merit, we compared BAMs generated from our proposed framework and its three variations. The first variation was trained only based on $H_-$, which means no non-referable DR data or the assistant generator were used. The second variation was trained based on $H_-$ and $L_c$, which means no preserved output was generated. The third variation was trained based on $H_-$ and $M_-$, which means no assistant generator was used. Except for the BAMs generated from our proposed framework, the BAMs of the three variations all highlighted features not related to DR pathology such as normal microvasculature and large vessels (marked by blue arrows in Fig. 10). In addition, the foveal avascular zone could not be distinguished from the pathological NPAs in the $BAM_{+/-}$ generated with these three training variations (Fig. 10).

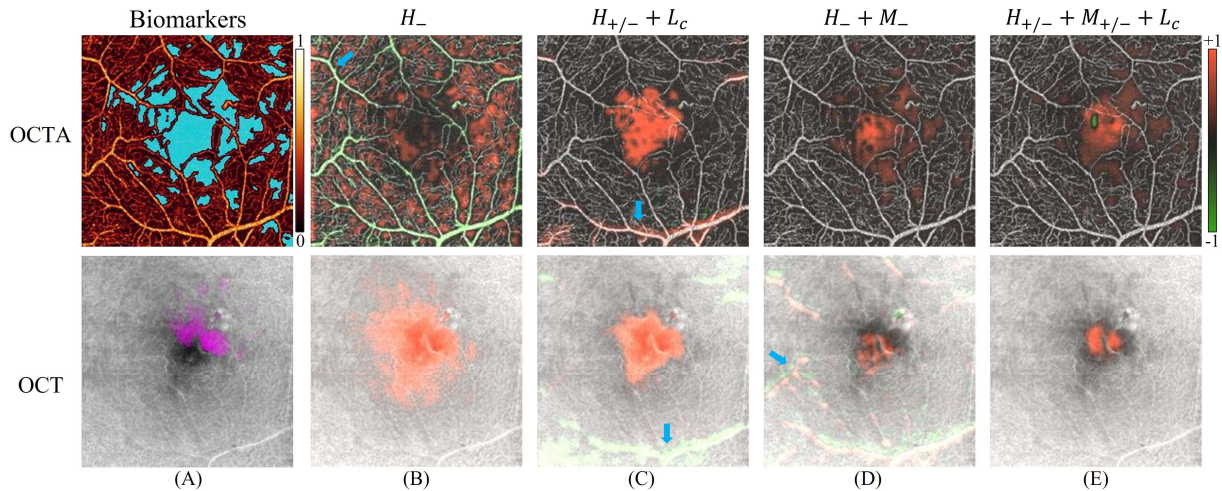

Fig. 10. BAMs generated in the ablation experiments. Large vessels highlighted by the three variations are marked by blue arrows. (A) Segmented non-perfusion areas and fluids. (B) The $BAM_{+/-}$ generated without non-referable DR data and assistant generator (loss: $H_-$). (C) The $BAM_{+/-}$ generated without preserved output (loss: $H_{+/-} + L_c$). (D) The $BAM_{+/-}$ generated without the assistant generator (loss: $H_- + M_-$). (E) The $BAM_{+/-}$ generated based on proposed framework (loss: $H_{+/-} + M_{+/-} + L_c$).

## VI. DISCUSSION

We proposed a BAM generation framework to aid in the interpretation of deep-learning-aided DR diagnosis. The core design concept of our framework is based on the recognition of unique requirements of DR diagnosis compared to non-medical image classification. By designing around this principle, we implemented a framework that enables visualization of specific biomarkers, rather than highlighting shared features between non-referable and referable OCT/OCTA images, which are irrelevant for verifying DR classifier outputs. The framework consists of two U-shaped generators (a main and an assistant generator), and was trained using generative adversarial learning. The BAMs clearly highlight biomarkers utilized by the classifier, which facilitate identification of clinically recognized pathology. The $BAM_{+/-}$ can also distinguish multiple features in the same image. To the best of our knowledge, the proposed BAM generation framework is the first interpretability method specifically designed for deep-learning-aided DR classifiers based on both OCT and OCTA. Based on both qualitative and quantitative comparisons between the BAM and attention maps generated by other methods our framework achieved state-of-the-art performance in providing interpretability to a DR classifier based on OCT and OCTA.

Existing interpretability methods were designed for non-medical image classification and produce attention maps that are not necessarily useful for validating classifier decision making in a medical context like DR diagnosis (Fig. 6). A lack of interpretability in medical deep-learning classifiers could lead to ethical and legal challenges, and as such a heuristic method that can provide sufficient interpretability for deep-learning classifiers is now recognized as an urgent need [21-23]. It is difficult to investigate bias if the reasons for the classifier's decisions are unclear [21, 22]. In part to address these concerns, the European Union's General Data Protection Regulation law requires that algorithm decision-making be transparent before it can be utilized for patient care [23, 86].

To evaluate the performance of our BAM, the DR classifier was forced to learn and utilize NPA and fluids which were two important pathologies (clinical biomarkers) for DR diagnosis. Among them, NPA is a biomarker closely correlated to ischemia which is a critical consequence that can be found in the early stage of DR [7, 75, 84]. Fluid is a biomarker closely correlated to DME which is the most common cause of vision loss in DR [7, 85]. In this study, the ground truth NPA and fluid were segmented by previously proposed deep learning methods, respectively [84, 85]. However, these two segmentation methods have no correlation with the interpretability; they were designed to segment all the areas with NPA/fluids no matter whether these clinical biomarkers areas were utilized by the classifier or not. (For example, the fluid segmentation method would identify fluid outside of the macula even though this is subclinical feature for DME.) Only our BAM could provide sufficient interpretability to a DR classifier by accurately highlight the classifier-utilized biomarkers. Clinicians could then verify whether these classifier-utilized biomarkers are clinical biomarkers or not.

In addition to the commonly used attention maps with only positive values ($BAM_{abs}$), we also generated $BAM_{+/-}$ which separating the positive and negative highlighted biomarkers. There are two major significances of generating $BAM_{+/-}$. Firstly, we could separate the biomarkers that were differently understood by the classifier. Especially the adjacent biomarkers like foveal avascular zone (marked by green in Fig. 6(E)) and surrounded pathological NPAs (marked by red in Fig. 6(E)), and hypo- and hyper- reflective spots (marked by red and green in Fig. 9(J), respectively). Secondly, with $BAM_{+/-}$, we could better understand how a biomarker was learned and utilized by the classifier. The highlighted areas in Fig. 6(E) shown the classifier has learned what the anatomical structure near fovea in a normal eye should be. The highlighted areas shown in Fig. 7(C) and 7(F) told us the classifier not only learned the intensity of these areas in a negative case (without rectangle artifacts) should be lower (much more green than red), but also learned the intensity in these areas of a negative case should not be even (have both green and red).

Generative adversarial learning has been used in several methods to generate pseudo-healthy (corresponding to forged negative in our study) images [87-90]. However, compared to our BAM framework, the different maps generated by these methods cannot be used to provide interpretability to a DR classifier. Firstly, because these methods were trained based on the ground truth labels using several discriminators, which means their generation results only correlated to the data set and could have no correlation with the classifier. Secondly, they only focused on the detection of each pathology. But a DR classifier may not utilize all the pathologies if only a subset of was sufficient for the diagnostic task. This appears to be the case with our classifier, which largely ignored pathology like hyper-reflective foci or microaneurysms. On the contrary, our framework was trained to only remove the classifier-utilized biomarkers from input positive images by using predicted labels without any discriminator. Therefore, compared to previously proposed pseudo-healthy image generation methods, only our framework could be used to provide adequate interpretability to a DR classifier.

Technically, the cycle-consistency generative adversarial learning strategy was used to train our framework [34]. However, compared to the original architecture and training protocol, our framework had several innovations. Firstly, the two discriminators were replaced by a DR classifier that will be interpreted. This design means the framework had the ability to learn provide a heuristic for interpreting the DR classifier. Secondly, the model selection was based on the loss function of only one generator (the main) since our goal was to highlight all the classifier-utilized biomarkers that only belong to referable DR. This design allowed us to select the main generator with highest performance. Thirdly, the generator output was calculated as the sum of input and Tanh output with zero initialization. This design avoided changes of the biomarkers which were not utilized by the classifier in the beginning of the training.

Though our BAM generation framework was only evaluated on an OCT/OCTA-based DR classifier, it can be easily transferred to interpretability tasks for other disease classifiers. For a binary disease classifier, the main generator always carries inputs to forged negative outputs, and vice versa for the assistant generator. For a single disease classifier which classifies each input to $S$ $(S \geq 3)$ severities, overall $S - 1$ BAMs are needed to provide sufficient interpretability to the whole classifier (not just the diagnosis of one severity). Each BAM is generated between two adjacent severities by respectively combing all lower and higher severities as one class. Alternatively, for a multiple disease classifier (e.g. a system that diagnoses DR and age-related macular degeneration) a BAM for each disease can be generated between the selected disease and normal class.

In addition to providing interpretability to disease classifiers, our BAM could also facilitate identification of new biomarkers and assessment of pharmacological impact via medical imaging. For example, if a disease classifier were trained on an imaging modality in which the disease is not well characterized the generated BAM could indicate features that should be explored. In drug development, for a classifier trained to classify the cases before and after the treatment, the generated BAM could highlight all the changes caused by the new drug.

Unlike other interpretability methods, our BAM generation framework uses deep learning networks to interpret another deep learning network, which leads to its own questions about interpretability. For networks designed for classification or segmentation, the interpretability issue can be described as how the classification or segmentation results are acquired. In medical image analysis, the concern for interpretability can be further described as whether the clinically meaningful biomarkers are used by the network to make decisions. However, the training target of our framework is generating an output which can be classified as negative by the trained classifier from a positive input. The interpretability issue in this context can be described as asking how the output is generated to achieve the desired classification, and asking which biomarkers correlate with classifier decision making. The BAM apparently improved interpretability by highlighting the biomarkers that were involved in changing the output of the trained classifier. Therefore, our BAM generation framework is self-interpreted and can be used to provide interpretability to deep-learning-aided disease classifiers.

Though our BAM could accurately highlight the classifier-utilized biomarkers and was sensitive to the potential interpretability changes, the anatomical structures of the forged outputs still look different from real data. In a future study, we will modify our BAM to not only highlight the classifier-utilized biomarkers but also generate the forged output with similar anatomical structure of the real data.

## VII. Conclusion

We proposed a BAM generation framework which can be used to provide interpretation of deep-learning-aided DR classifier. The BAMs demonstrated here accurately highlighted different classifier-utilized biomarkers at high resolution, which enable quick review by image graders to verify whether clinically meaningful biomarkers were used by the classifier. Our BAM generation framework can improve the clinical acceptability and real-world applications for deep-learning-aided DR classifiers.

## Acknowledgment

Oregon Health & Science University (OHSU), Yali Jia and David Huang have a significant financial interest in Visionix, Inc. Yali Jia has a significant financial interest in Optos, Inc. These potential conflicts of interest have been reviewed and managed by OHSU.